\documentclass[twocolumn]{aastex63}
\usepackage{epsfig}
\usepackage{amsmath, amsthm, amssymb}
\usepackage{microtype}
\usepackage{float}
\usepackage{verbatim}
\usepackage{wrapfig}
\usepackage{graphicx}
\usepackage[caption = false]{subfig}
\usepackage{multirow}
\usepackage{hhline}
\usepackage{booktabs}
\usepackage{array}
\usepackage{bm}
\usepackage[normalem]{ulem}

\shorttitle{A new candidate redback MSP}
\shortauthors{Swihart et al.}

\begin{document}

\title{Discovery of a New Redback Millisecond Pulsar Candidate: 4FGL J0940.3--7610}

\correspondingauthor{Samuel J. Swihart}
\email{samuel.swihart.ctr@nrl.navy.mil}

\author{Samuel J. Swihart}
\affiliation{National Research Council Research Associate, National Academy of Sciences, Washington, DC 20001, USA,\\ resident at Naval Research Laboratory, Washington, DC 20375, USA}
\affiliation{Center for Data Intensive and Time Domain Astronomy, Department of Physics and Astronomy,\\ Michigan State University, East Lansing, MI 48824, USA}
\author{Jay Strader}
\affiliation{Center for Data Intensive and Time Domain Astronomy, Department of Physics and Astronomy,\\ Michigan State University, East Lansing, MI 48824, USA}
\author{Elias Aydi}
\affiliation{Center for Data Intensive and Time Domain Astronomy, Department of Physics and Astronomy,\\ Michigan State University, East Lansing, MI 48824, USA}
\author{Laura Chomiuk}
\affiliation{Center for Data Intensive and Time Domain Astronomy, Department of Physics and Astronomy,\\ Michigan State University, East Lansing, MI 48824, USA}
\author{Kristen C. Dage}
\affiliation{Center for Data Intensive and Time Domain Astronomy, Department of Physics and Astronomy,\\ Michigan State University, East Lansing, MI 48824, USA}
\author{Laura Shishkovsky}
\affiliation{Center for Data Intensive and Time Domain Astronomy, Department of Physics and Astronomy,\\ Michigan State University, East Lansing, MI 48824, USA}

\begin{abstract}

We have discovered a new candidate redback millisecond pulsar binary near the center of the error ellipse of the bright unassociated \emph{Fermi}-LAT $\gamma$-ray source 4FGL J0940.3--7610. The candidate counterpart is a variable optical source that also shows faint X-ray emission. Optical photometric and spectroscopic monitoring with the SOAR telescope indicates the companion is a low-mass star in a 6.5-hr orbit around an invisible primary, showing both ellipsoidal variations and irradiation and consistent
with the properties of known redback millisecond pulsar binaries. Given the orbital parameters, preliminary modeling of the optical light curves suggests an edge-on inclination and a low-mass ($\sim 1.2$--$1.4\,M_{\odot}$) neutron star, along with a secondary mass somewhat more massive than typical $\gtrsim 0.4\,M_{\odot}$. This combination of inclination and secondary properties could make radio eclipses more likely for this system, explaining its previous non-discovery in radio pulsation searches. Hence 4FGL J0940.3--7610 may be a strong candidate for a focused search for $\gamma$-ray pulsations to enable the future detection of a millisecond pulsar.
\end{abstract}

\vspace{5mm}
\section{Introduction}
Pulsars form the largest population of Galactic \emph{Fermi}-LAT $\gamma$-ray sources with clear associations \citep{4FGLcatalog}. Follow-up studies of as-yet unidentified $\gamma$-ray sources continue to reveal new compact binaries, especially those containing millisecond pulsars (MSPs) spun up to fast periods through accretion from a companion \citep[e.g.,][]{Parent19,Corongiu20,Wang20}. $\gamma$-ray emission from these objects may be ubiquitous \citep{2PCfermi} and can serve as a signpost for MSPs difficult to find or study at other wavelengths, such as ``spider" (black widow or redback) MSP binaries with non-degenerate companions that have extensive eclipses in the radio \citep[e.g.,][]{Polzin19,Crowter20, Kudale20, Pan20}.

Finding the multi-wavelength (optical, X-ray, or radio) counterpart of unidentified \emph{Fermi}-LAT $\gamma$-ray sources can be challenging. The $\gamma$-ray error ellipses are relatively large (often dozens of square arcminutes on the sky), potentially containing a substantial number of X-ray or radio sources and hundreds or more optical sources.

Characteristic multi-wavelength behavior can help narrow down possible associations for spider MSPs. Most of these show a characteristic hard X-ray spectrum with rapid stochastic and orbital variability, likely due to an intrabinary shock that occurs between the wind from the companion star and the pulsar wind \citep[e.g.,][]{Gentile14,Wadiasingh17,Romani16,AlNoori18}. In addition, $\gamma$-ray emitting compact binaries with secondaries that substantially fill their Roche lobes should have detectable periodic optical variability due to the tidal deformation (and sometimes irradiation) of the secondary star \citep[e.g.,][]{Romani11,Strader15,Halpern17,Li18,Swihart20}.

In this paper, we present the discovery of an X-ray and variable optical source that we argue is likely a new redback MSP binary associated with the unidentified \emph{Fermi} source 4FGL J0940.3--7610.

\begin{figure*}[ht!]
\begin{center}
	\includegraphics[width=0.48\linewidth]{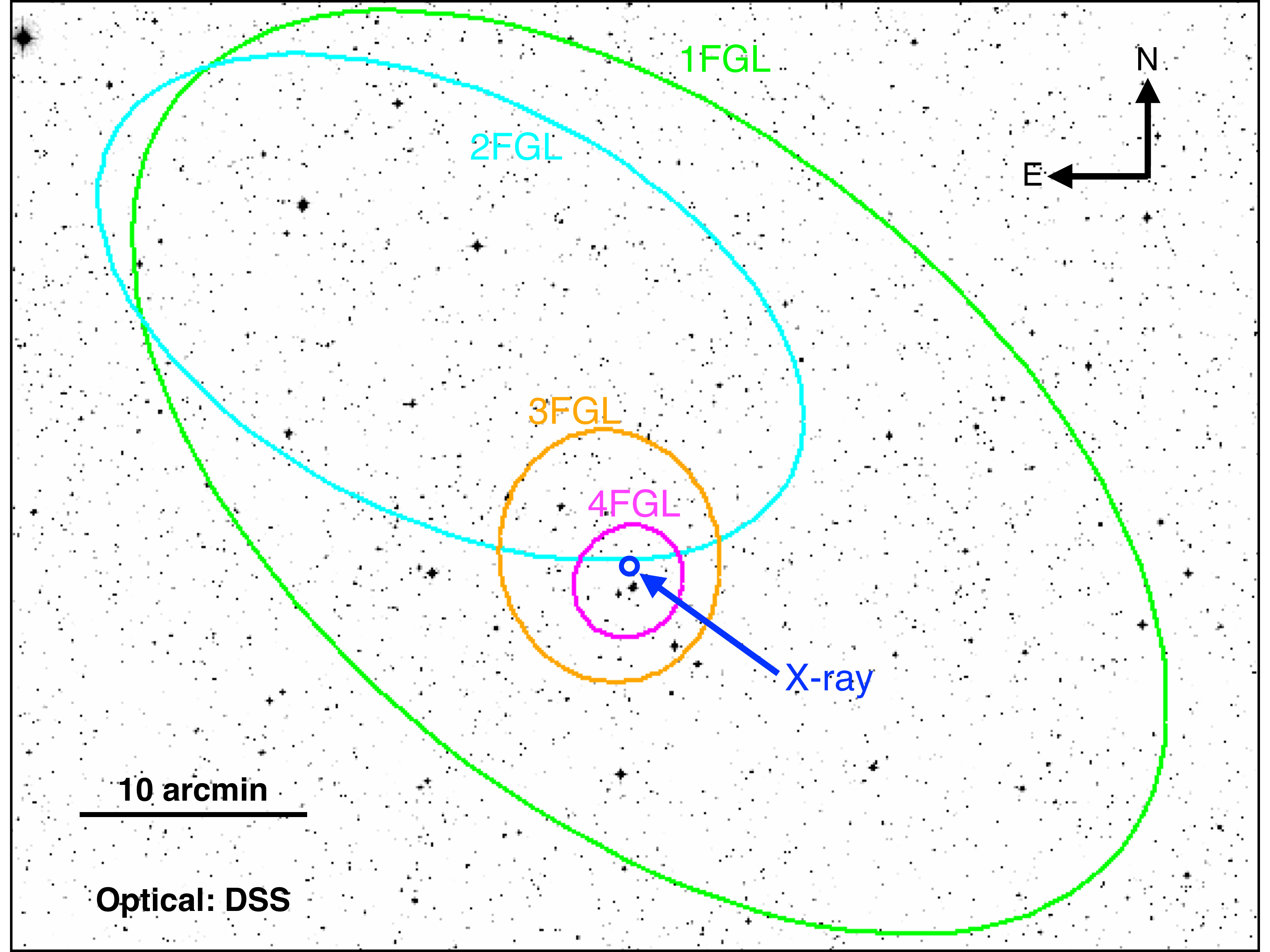}
	\hspace{5mm}
	\includegraphics[width=0.48\linewidth]{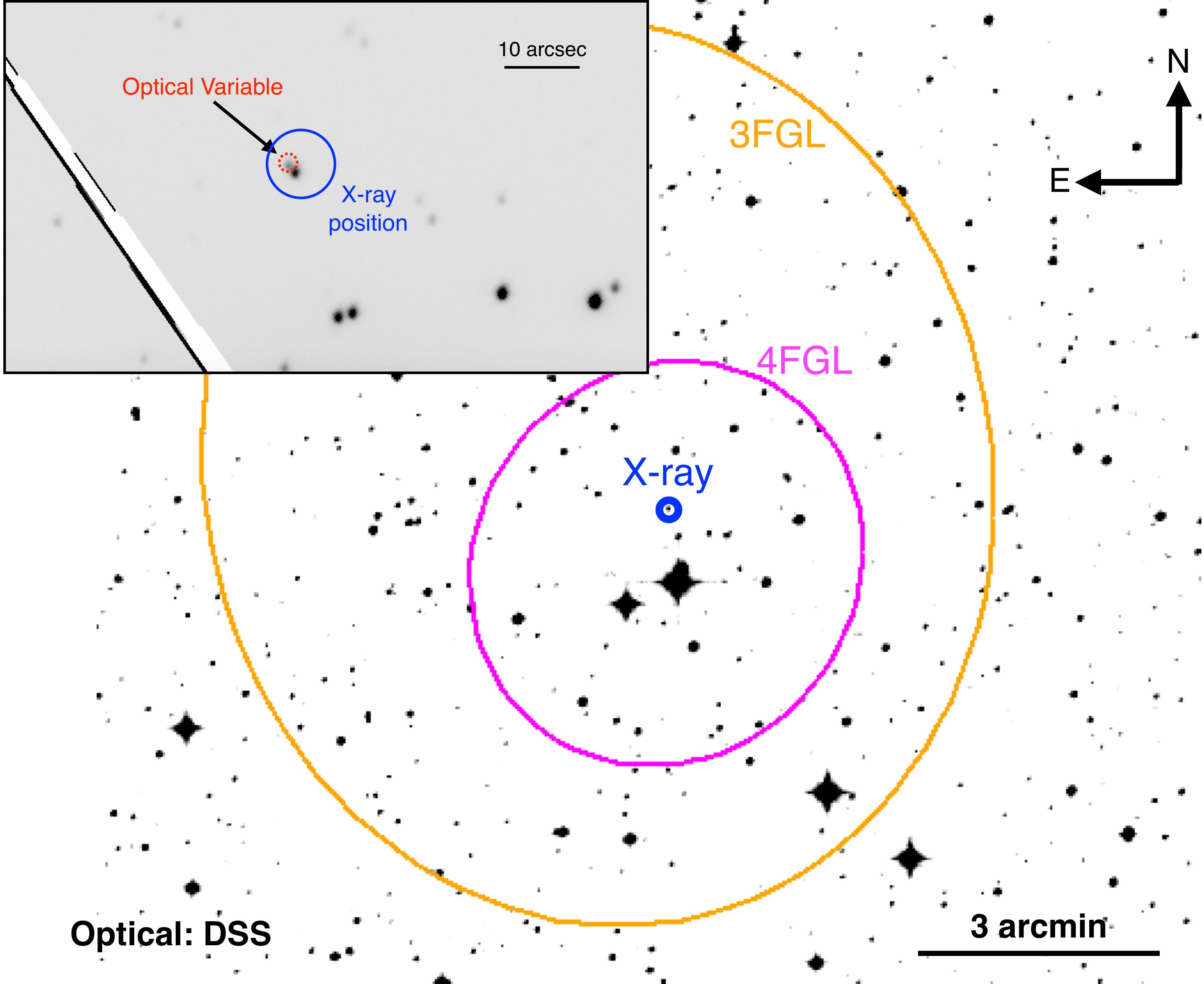}
    \caption{Left: Optical Digitized Sky Survey image of the field showing the positions and overlapping 95\% error ellipses from the 1FGL, 2FGL, and 3FGL catalogs corresponding to the $\gamma$-ray source 4FGL J0940.3--7610 (magenta), along with the position of the \emph{Swift} X-ray source (blue circle, see \S~\ref{sec:X-rays}). Right: Same as the left figure now zoomed-in on the 3FGL region. The inset displays a zoomed-in view of one of our SOAR $i'$ images along with the \emph{Swift} X-ray position and 90\% confidence region (blue). The optical variable that is the subject of this work is marked with a red circle, and is $\sim$1.4\arcsec~away from a nearby, brighter star that is unrelated to the $\gamma$-ray source. The large streak in the inset is a diffraction spike from a nearby bright star.}
\end{center}
\label{fig:finder_fig}
\end{figure*}

\section{Observations}

\subsection{The $\gamma$-ray Source}
\label{sec:gammarays}
The $\gamma$-ray source that is the subject of this work was reported in the first full catalog of \emph{Fermi}-LAT sources, based on the first eleven months of survey data, and was listed as 1FGL J0940.2--7605 \citep{1FGLcatalog}. It has appeared in each subsequent catalog, most recently as 4FGL J0940.3--7610 in the LAT ten-year source catalog \citep{4FGLDR2}. The 95\% error ellipse is nearly circular and relatively small compared to most unassociated LAT sources, with a mean radius of $\sim$2.6\arcmin. The $\gamma$-ray source shows a spectrum with significant curvature ($6.6\sigma$ for the LogParabola model; $7.1\sigma$ for a cut-off power law). The 0.1--100 GeV energy flux is $(6.5\pm0.6) \times 10^{-12}$ erg s$^{-1}$ cm$^{-2}$, which is a strong detection as expected for a source present from the 1FGL catalog. There is no significant evidence for variability \citep{4FGLDR2}.

The spectral curvature and lack of evidence for variability are typical of $\gamma$-ray observations of pulsars \citep{2PCfermi}. Indeed, this source is listed as a likely MSP in several papers that used machine learning to classify unassociated LAT sources from 3FGL or 4FGL \citep[e.g.,][]{Saz16,Luo20}.

\subsection{X-rays}
\label{sec:X-rays}
The field of 4FGL J0940.3--7610 was observed with the Swift X-ray Telescope (XRT) nine times between 2011 and 2015 as part of a program following-up unidentified \emph{Fermi}-LAT sources \citep{Stroh13}. The total good exposure time on source is $\sim$2.3 ksec.

In these data, there are two X-ray sources within the updated 4FGL error ellipse listed in the 2SXPS catalog \citep{Evans20}, one significant and one marginal. The X-ray source that is the subject of this paper, with a catalog ID of 2SXPS J094023.5--761001, is both brighter and closer to the center of the $\gamma$-ray error ellipse, only 
45\arcsec~from its center, with an ICRS (R.A., Dec.) of (09:40:23.54, --76:10:01.0) and a 90\% positional uncertainty of 4.6\arcsec~(Figure~\ref{fig:finder_fig}). The more marginal X-ray source is near the edge of the 4FGL error ellipse and has no apparent optical counterpart.

2SXPS J094023.5--761001 has a 0.3--10 keV count rate of $(6.8\pm2.1) \times 10^{-3}$ ct s$^{-1}$. Assuming a power law spectrum with standard redback photon index of $\Gamma = 1.4$ and a foreground $N_H = 9.5 \times 10^{20}$ cm$^{-2}$ \citep{HI4PI16} this corresponds to an unabsorbed flux of $(3.8\pm1.2) \times 10^{-13}$ erg s$^{-1}$ cm$^{-2}$. The photon index inferred from the hardness ratios is $\Gamma = 0.9^{+1.5}_{-1.0}$, which is nominally hard but with large enough uncertainties to admit nearly any interpretation.

\subsection{Optical Spectroscopy}
There are two optical sources in \emph{Gaia} DR2 \citep{GaiaDR2}
within the astrometric uncertainty of 2SXPS J094023.5--761001. These optical sources are separated by only 1.4\arcsec~(Figure~\ref{fig:finder_fig}) and so in principle either could be the counterpart to the X-ray source. We note that these two \emph{Gaia} sources have very different proper motions and hence are not comoving.

We obtained optical spectroscopy of both these targets using the Goodman spectrograph \citep{Clemens04} on a two-night run with the SOAR telescope on 2018 Dec 10/11. The brighter of these two sources ($G=18.2$) showed no evidence for radial velocity variations among different epochs, while the fainter source ($G=19.3$) evinced large (several hundred km s$^{-1}$) shifts over just a few hr on 2018 Dec 11, proving that it is a close binary. Hence we identified this fainter optical source as the most likely counterpart to the X-ray source. The \emph{Gaia} DR2 ICRS position of this source is (R.A., Dec.) = (09:40:23.787, --76:10:00.13), which we take as the best position available. 

We performed spectroscopic monitoring of this source 
over seven nights from 2018 Dec 11 to 2019 Mar 25. For each spectrum we used a 400 l mm$^{-1}$ grating and a 0.95\arcsec~slit, yielding a resolution of about 5.6 \AA\ (full width at half-maximum). All spectra covered a wavelength range of $\sim 4800$--8800 \AA. Exposure times were either 20 to 25 min per spectrum, depending on conditions. The slit was oriented at a parallactic angle of 134.6$^{\circ}$ to avoid contamination from the nearby brighter source.

Each spectrum was reduced and optimally extracted in the normal manner using standard packages in \texttt{IRAF} \citep{Tody86}. We measured barycentric radial velocities through cross-correlation with bright template stars of similar spectral type, with the cross-correlations done simultaneously in a bluer region around Mg$b$ and a redder region from 
$\sim 6050$--6360~\AA. The resulting 35 radial velocities are listed in Table~\ref{table:RVdata}.
The observation times, representing the mid-point of each observation, are reported as Modified Barycentric Julian Dates (BJD -- 2400000.5 d) on the TDB system \citep{Eastman10}.

\begin{deluxetable}{crr}[h]
\label{table:RVdata}
\tablecaption{Modified Barycentric Radial Velocities of 4FGL J0940.3--7610}
\tablehead{MBJD & RV & err. \\
                   (d)  & (km s$^{-1}$) & (km s$^{-1}$)}
\startdata
58463.2613142 & 322.2 & 22.9 \\
58463.2787955 & 334.6 & 24.4 \\
58463.2998051 & 184.1 & 23.1 \\
58463.3174170 & 125.2 & 24.4 \\
58484.1968846 & --158.4 & 23.1 \\
58484.2147938 & --227.9 & 27.4 \\
58484.2360225 & --291.2 & 21.9 \\
58484.2534997 & --266.3 & 24.8 \\
58484.2922107 & --41.4 & 26.1 \\
58491.2160776 & --65.7 & 24.7 \\
58491.2336100 & --178.8 & 25.2 \\
58491.2564964 & --193.9 & 22.8 \\
58491.2957546 & --166.2 & 21.9 \\
58491.3132683 & --154.9 & 22.9 \\
58525.0872665 & --270.6 & 25.0 \\
58525.1047505 & --297.7 & 25.3 \\
58525.1238465 & --218.9 & 23.8 \\
58525.1378938 & --151.2 & 24.2 \\
58525.1563986 & --32.0 & 22.3 \\
58525.1742462 & 39.1 & 20.7 \\
58525.2992549 & 94.6 & 24.1 \\
58525.3167359 & --25.6 & 20.9 \\
58537.2547892 & --237.7 & 24.5 \\
58537.2725452 & --268.0 & 22.1 \\
58547.1325566 & 254.7 & 24.5 \\
58547.1506098 & 301.1 & 21.8 \\
58547.1730925 & 302.8 & 27.6 \\
58547.1905521 & 252.7 & 23.0 \\
58547.2080537 & 129.4 & 24.9 \\
58567.1326788 & 117.4 & 27.4 \\
58567.1501386 & 216.6 & 26.8 \\
58567.1896342 & 338.9 & 25.7 \\
58567.2305353 & 175.5 & 29.3 \\
58567.2527752 & 19.6 & 24.8 \\
58567.2702688 & --67.7 & 22.0
\enddata
\label{table:RVdata}
\end{deluxetable}

\subsection{Optical Photometry}

To obtain light curves of the optical source, we also performed imaging observations with SOAR/Goodman on two nights in 2019, with the CCD binned 2x2 to a pixel scale of 0.3\arcsec\ per pixel. On 2019 Apr 08 we used a series of alternating exposures with the $g'$ (exposure time 300 s) and $i'$ (180 or 240 s) filters, with median seeing of about 1.5\arcsec. On 2019 Apr 21 we observed only in $i'$ (300 s per exposure), and the median seeing was about 1.2\arcsec.

The reduction was done as outlined in \citet{Swihart20}, including bias correction and flat fielding. We performed differential aperture photometry with respect to nearby, non-variable comparison stars, using 36 stars in $g'$ and 32 stars in $i'$. We then calibrated these magnitudes to a standard system, using \emph{Gaia} DR2 $G$ mag for $g'$ \footnote{https://gea.esac.esa.int/archive/documentation/GDR2/ Data\_processing/chap\_cu5pho/sec\_cu5pho\_calibr/
ssec\_cu5pho\_PhotTransf.html} and Skymapper for $i'$ \citep{Wolf18}. Finally, we made a frame-by-frame correction to account for the ``excess" flux from the non-variable nearby (1.4\arcsec~in projection) source, by measuring its magnitude in both a small aperture and the combined magnitude of both sources in a large (10\arcsec) aperture.

Our final sample consists of 36 and 102 photometric measurements in $g'$ and $i'$, respectively. The mean observed magnitudes of our target are $g' = 19.754$ mag and $i' = 18.832$ mag, with median uncertainties (statistical + systematic) of $\sim$0.060 and $\sim$0.035 mag in $g'$ and $i'$, respectively.\\

\section{Results and Analysis}

\subsection{Optical Spectroscopy and Orbital Parameters}
\label{sec:spec_results}
Overall, the spectra of the optical source are consistent with that of a moderately cool, late-G to early-K type dwarf star (we reach similar conclusions from our light curve fitting in \S~\ref{sec:LCmodeling}). This spectral type is typical of redback secondaries \citep[e.g.,][]{Bellm16, Linares18b, Swihart19, Swihart20}. There is no evidence for a second set of absorption lines in any of the optical spectra, as would be expected if this star was in a binary with a main sequence or evolved star, suggesting the companion is a compact object. We see no emission lines in any of our spectra, indicating that an accretion disk is not present; Balmer emission has also been observed in some non-accreting redbacks, and is likely due to the presence of an ionized wind or shock \citep{Swihart18}.

In order to determine the orbital properties of the system, we fit Keplerian models to the radial velocities using the custom Monte Carlo sampler \texttt{TheJoker} \citep{Price17}. We initially fit for the binary period $P$, time of ascending node $T_0$, orbital semi-amplitude $K_{2}$, systemic velocity $\gamma$, and the eccentricity $e$.  However, we found no significant evidence for non-zero eccentricity, so for the remainder of our analysis we assume a circular orbit. This is expected, since in the absence of an additional perturbing body, the binary is expected to be circularized in a very short timescale ($\lesssim 10^4$ yr) at the observed orbital period for a typical redback mass ratio \citep{Zahn77}.

\begin{figure}[t]
\includegraphics[width=1.0\linewidth]{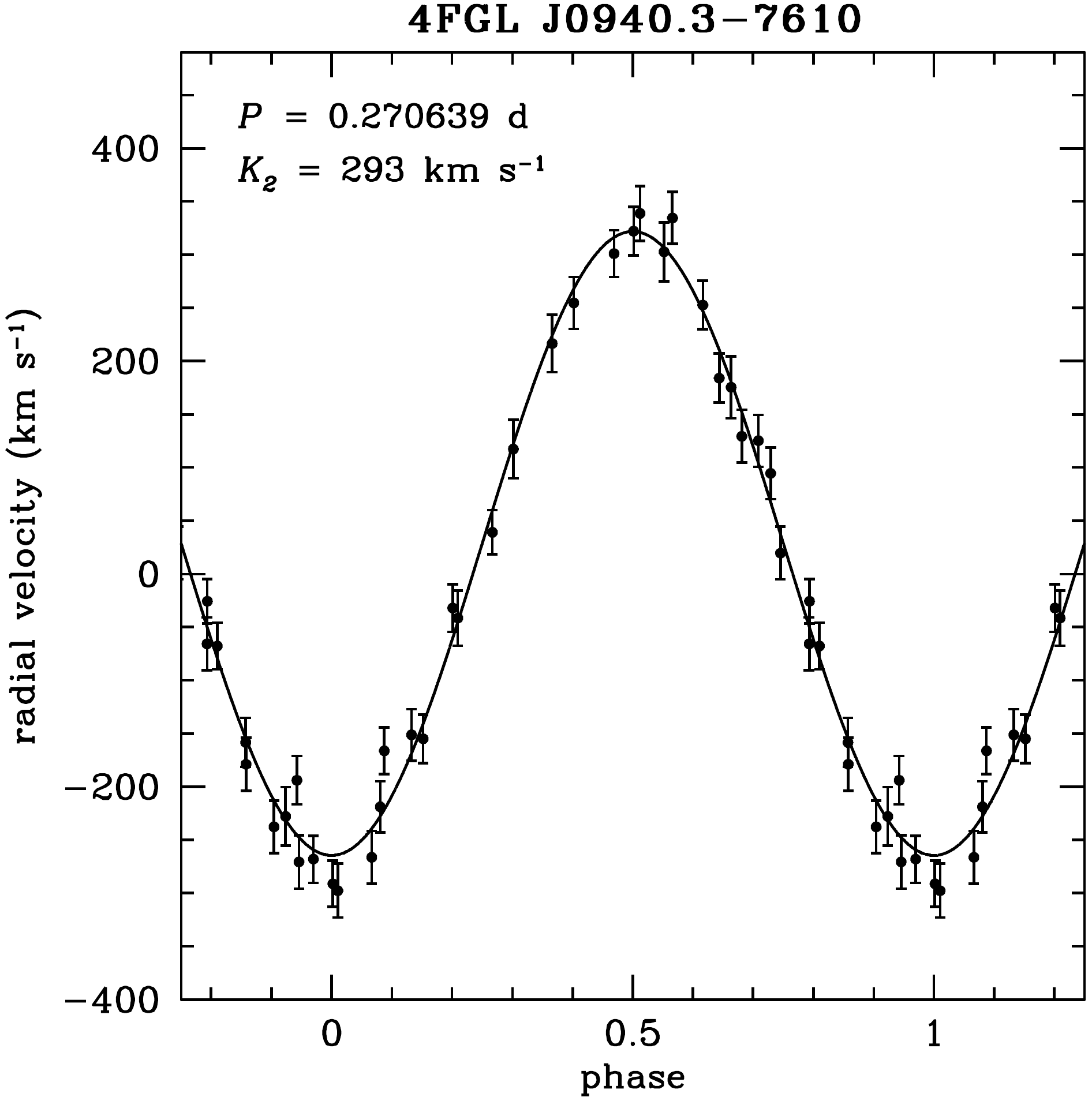}
\caption{Circular Keplerian fit to the SOAR/Goodman barycentric radial velocities of the optical counterpart to 4FGL J0940.3--7610.}
\label{fig:rvmodel}
\end{figure}

For our circular model, we find $P = 0.270639(7)$ d, $T_{0} = 58525.37263(172)$ d, semi-amplitude $K_{2} = 293.2 \pm 6.0$ km s$^{-1}$, and $\gamma = 28.7 \pm 4.1$ km s$^{-1}$. Overall the fit is very good statistically, with a rms of 22.6 km s$^{-1}$ and a $\chi^2$/d.o.f. = 33.2/31. We plot this model along with the radial velocity data in Figure~\ref{fig:rvmodel}.

We derive the binary mass function $f(M)$ using posterior samples from our radial velocity modeling:

\begin{equation}
    f(M) = \frac{P \, K_{2}^{3} \, (1-e^2)^{3/2}}{2\pi G} = \frac{M_{1} \, \textrm{sin}^3 i} {(1+q)^{2}},
\end{equation}

\noindent
where $M_{1}$ is the mass of the primary, $q = M_{2}/M_{1}$ is the mass ratio, and $i$ is the system inclination. Using our spectroscopic results, $f(M) = 0.71 \pm 0.04\,M_{\odot}$, which represents a lower limit on the primary mass. Any normal star of this minimum mass would be apparent in the optical spectra; since it is not seen, the clear implication is that the primary is a compact object. We note that this value of the mass function is fairly typical for redbacks \citep{Strader19}. 

Neither $q$ nor $i$ can be well-constrained from our low-resolution optical spectroscopy alone, but some information can be obtained from modeling the light curves, which we do in the next subsection.

\subsection{Light Curve Modeling}
\label{sec:LCmodeling}
\begin{figure}[t]
	\includegraphics[width=\linewidth]{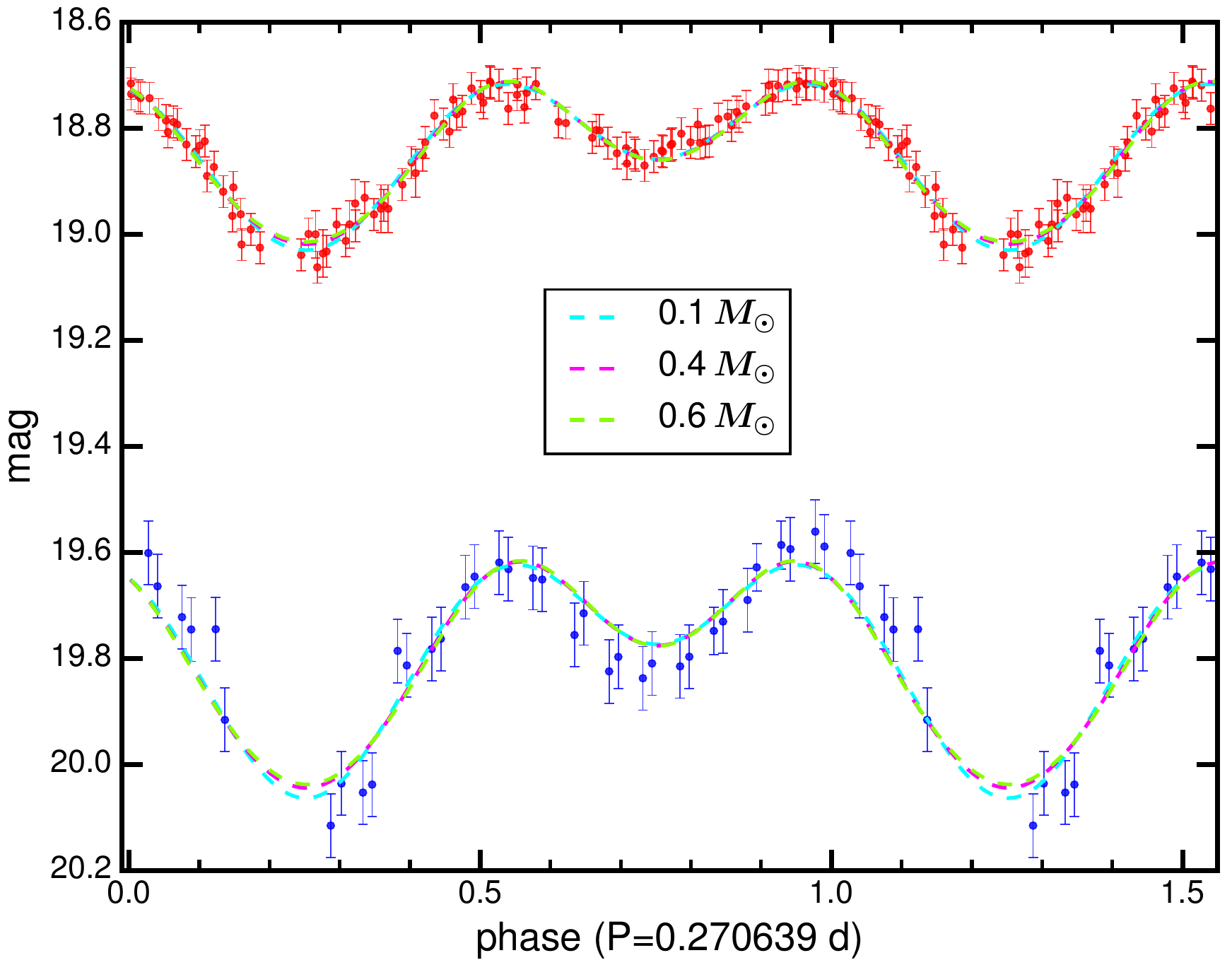}
    \caption{SOAR $g'$(blue) and $i'$(red) photometry of the optical counterpart to 4FGL J0940.3--7610 folded on the best-fit period and ephemeris from our radial velocity measurements. The best-fit ELC light curve models with varying secondary masses are shown with dashed lines. The differences between the models are small, but only models with a secondary mass $\gtrsim 0.35\,M_{\odot}$ can accommodate a nuetron star-mass primary.}
\label{fig:lcmodel}
\end{figure}

Folding the $g'$ and $i'$ photometry on the spectroscopic orbital period shows the characteristic double-peaked morphology expected for a tidally-distorted secondary,
with two equally bright maxima when the system is viewed at quadrature (Figure~\ref{fig:lcmodel}). For simple ellipsoidal variations the expectation is that, due to gravity darkening, the secondary should be fainter at $\phi=0.75$ (at its superior conjunction) than at $\phi=0.25$. For 4FGL J0940.3--7610 we see the opposite, implying that the secondary is heated on its tidally-locked ``dayside''. For redbacks, this heating is typically inferred to be either directly from the pulsar wind or more indirectly by reprocessed emission from an intrabinary shock \citep{Romani16, Sanchez17, Wadiasingh18}.

We modeled our photometry using the Eclipsing Light Curve \citep[ELC,][]{Orosz00} code, assuming a compact invisible primary with no accretion disk, and a tidally distorted secondary with the orbital properties derived from our spectroscopic results.

Since the mass ratio is unknown and this parameter is very challenging to tightly constrain with a modest amount of optical photometry, we instead chose to do a series of fits with a broad range of typical redback companion masses ($\sim 0.1-0.6\,M_{\odot}$) and report the results for different assumed companion masses, requiring that the observed $K_2$ be matched. For each assumed companion mass we ran models fitting for the system inclination $i$, the Roche lobe filling factor $f_{2}$, the effective temperature $T_{2}$ of the companion, and the irradiating luminosity.

We note that ELC models the irradiation in a relatively simple manner, and for some other redbacks more complex heating models have been explored. These take into account reprocessed or magnetically-ducted emission from an extended intrabinary shock between the pulsar and companion, or redistibution of heat on the stellar surface due to diffusion and convection within the photosphere \citep[e.g.,][]{Romani16,Sanchez17,Voisin20}. Given the poorly constrained mass ratio of the binary and the modest amount of our photometry, for the purpose of this discovery paper we stick to these simpler models, while emphasizing the need for more sophisticated modeling once better data are available.

Now to the light curve modeling results. First, in all cases, we find the best fits come from models that are relatively edge-on, with inclinations $\gtrsim 70^{\circ}$. In all the models that were good statistical fits to the data, the effective temperature of the companion was $\sim 5050 \pm 200$ K, consistent with the spectral types inferred from the optical spectra. The Roche lobe filling factors were relatively well-constrained, with values ranging from $f_{2}\sim0.84-0.95$, corresponding to stellar radii of $R\sim0.35-0.67\,R_{\odot}$, depending on the model (the more massive companions are also larger). As revealed by the light curves, ellipsoidal variations from the partially Roche lobe-filling secondary are apparent, suggesting the level of heating is not as extreme as observed in some redbacks \citep[e.g.,][]{Schroeder14,Cho18,Linares18,Swihart20}. Our modeling supports this: we find the maximum ``dayside'' temperature of the secondary to be only $\sim$5400 K.

It is also immediately clear that models with companion masses on the lower-mass end of the typical redback mass distribution are inconsistent with the interpretation of the primary as a neutron star. For instance, setting the secondary mass equal to $0.1\,M_{\odot}$, the lowest known for a redback \citep[PSR J1622--0315;][]{Strader19} yields a best-fit model with $i\sim81^{\circ}$ and a resulting primary mass of $\sim0.9\,M_{\odot}$, too low for a neutron star. Such a system could in principle have a white dwarf primary, although X-ray emission would be difficult to explain since there is no evidence of accretion (see \S~\ref{sec:spec_results}).

For models with typical redback secondary masses (or even atypically high ones)
the models do agree with a neutron star interpretation. When setting the companion mass equal to the approximate median redback companion mass \citep[$0.4\,M_{\odot}$;][]{Strader19}, the best fit inclination is $81.9^{+7.9}_{-12.1}$, corresponding to a primary mass of $1.26^{+0.15}_{-0.03}\,M_{\odot}$.

If the companion is on the more massive end of the redback distribution, approaching $0.6\,M_{\odot}$, such as the confirmed redback PSR J1306--40 \citep{Keane17,Linares18,Swihart19} or the candidate redback 3FGL J0212.1+5320 \citep{Li16,Shahbaz17}, then the best fit inclination is $i=87.7^{+2.3}_{-13.9}$ corresponding to $M_{1} = 1.43^{+0.11}_{-0.01}\,M_{\odot}$.

These models are very good fits to the data and are nearly identical statistically, with $\chi^2$/d.o.f. = 79.1/133 and 81.2/133 for the models with $0.4\,M_{\odot}$ and $0.6\,M_{\odot}$ secondaries, respectively.
We show these models with the data, along with a model for a $0.1\,M_{\odot}$ secondary in Figure~\ref{fig:lcmodel}, where it is clear that variations in the secondary mass results in only small differences between the models.

The models perform most poorly in $g'$ near companion superior conjunction ($\phi=0.75$), which suggests our treatment of the heating may not fully describe the true heating in the system. As mentioned above, it is likely that the heating prescription is more complex than our simple point source model, such as from reprocessed emission from an extended intrabinary shock or diffusion and convection within the stellar photosphere \citep[e.g.,][]{Romani16, Voisin20}.

The next largest excursions from the model appear in the $g'$ data just after quadrature, near $\phi=1.0$, where the data appear brighter than expected from these models. Given the evidence for irradiative heating and the likely strong magnetic fields induced on the tidally-locked (and thereby rapidly-rotating) companion’s surface, it is possible that accelerated intrabinary shock particles are magnetically-ducted to the surface, mimicking one or more hot-spots \citep[e.g.,][]{Sanchez17}. These spots are one way to cause irregular heating patterns like these in the light curves of redback secondaries if the spots are shifted in azimuth to be off the line connecting the primary and secondary \citep[e.g.,][]{Strader19, Swihart19}. We do not attempt to model these spots in this discovery paper, but we suggest these more complex heating models be explored when future data become available.

Overall we conclude that, given the spectroscopic orbital parameters, the light curves suggest an edge-on orientation with a relatively low neutron star mass ($\sim1.2$--$1.4\,M_{\odot}$).

\subsection{Distance}

The optical binary has a marginally significant parallax listed in \emph{Gaia} DR2: $\varpi = 0.667\pm0.263$ mas \citep{GaiaDR2}. Taking into account a global parallax offset of +0.029 mas \citep{Lindegren18}, and using a simple exponential length prior of 1.35 kpc \citep{Astraatmadja16}, this parallax implies a distance of $1.6^{+1.6}_{-0.5}$ kpc.

We independently estimate the distance using our light curve models following the procedures described in \citet{Swihart17}. For the model with a $0.4\,M_{\odot}$ secondary (typical for a redback), the inferred distance is $2.2^{+0.5}_{-0.3}$ kpc, where the uncertainties represent the range of distances found for different assumptions for the dust reddening and metallicity \citep[see e.g.,][]{Swihart19}. Models with less (more) massive secondaries would give smaller (larger) distances; for example, a $0.1\,M_{\odot}$ secondary would give a distance in the range 1.4--1.7 kpc.

At this point, both the light curve modeling and \emph{Gaia} distances have substantial random and systemtic uncertainties, but also concur on a distance around $\sim 2$ kpc, which we take as a fiducial distance for the remainder of the paper. The precision of the \emph{Gaia} parallax distance should improve in future data releases.

\subsection{The X-ray and $\gamma$-ray emission}

The inferred 0.5--10 keV X-ray luminosity of 4FGL J0940.3--7610 is $(1.8\pm0.6) \times 10^{32} (d/2 \textrm{kpc})^2$ erg s$^{-1}$, a value squarely in the range for that observed for known redbacks in the pulsar state \citep[e.g.,][]{Linares14,Hui19,Strader19}. This is likewise the case for the 0.1--100 GeV $\gamma$-ray luminosity of $(3.1\pm0.3) \times 10^{33} (d/2 \textrm{kpc})^2$ erg s$^{-1}$, and as noted in \S~\ref{sec:gammarays} the $\gamma$-ray spectrum and lack of variability is consistent with the properties of MSPs.

\citet{Miller20} discuss the use of the ratio of X-ray to $\gamma$-ray flux ($F_X/F_{\gamma}$) to classify Galactic compact binaries, focusing on the distinction between spider MSPs and transitional MSPs in the sub-luminous disk state in a distance-independent manner. The median value of $F_X/F_{\gamma}$ for redbacks is 0.012, with a range from 0.003 to 0.12. For 4FGL J0940.3--7610, $F_X/F_{\gamma} = 0.06\pm0.02$, well within the distribution of other redback MSPs, and much lower than the values observed for candidate and confirmed transitional MSPs in the disk state (which have $F_X/F_{\gamma}$ in the range 0.28--0.43). Hence $F_X/F_{\gamma}$ is consistent with the classification of 4FGL J0940.3--7610 as a normal redback in the pulsar state.

The properties of the binary are not well-explained as a chance alignment with a $\gamma$-ray source: while the orbital properties and light curve modeling could admit a solution with a primary of white dwarf mass, there is no evidence in the optical spectra for accretion, without which it would be hard to reach the observed $L_X \sim 10^{32}$ erg s$^{-1}$, or to produce the observed irradiation in the light curves.

\section{Discussion and Conclusions}
\label{sec:discussion}

We have discovered a short-period (6.5 hr) compact binary with X-ray emission near the center of the error ellipse of the unassociated \emph{Fermi} $\gamma$-ray source 4FGL J0940.3--7610. The optical and X-ray properties of the source are well-explained as a redback MSP but are not consistent in a straigtforward manner with any other class of source. Hence we think the binary is likely to be a redback millisecond pulsar and the counterpart to 4FGL J0940.3--7610.

Compared to known (or strong candidate) redbacks, the orbital period, high-energy properties, and distance are typical. The inclination appears relatively edge-on. The mass of the neutron star appears to be low, closer to $\sim 1.4\,M_{\odot}$ as for the redbacks PSR J1723–2837 \citep{vanStaden16} and PSR J2039--5617 \citep{Clarke20} rather than $\sim 1.8\,M_{\odot}$ as found for a typical (median) redback \citep{Strader19}.

There is also a hint that the secondary might be on the more massive side for redback companions (perhaps $\gtrsim 0.4\,M_{\odot}$). This has potential relevance for the fact that a radio pulsar has not yet been detected toward this region despite extensive searches \citep[e.g.,][]{Camilo16}. There are several other systems that have compelling optical and X-ray evidence for being redbacks but in which pulsars have also not been found, such as
1FGL J0523.5–2529 \citep{Strader14} and 3FGL J0212.1+5320 \citep{Li16, Linares17, Shahbaz17}, which have secondaries that are more massive than typical for redbacks. This could suggest that the pulsars in these systems are more difficult to detect through radio observations, perhaps due to more extensive eclipses. This makes 4FGL J0940.3--7610 a strong candidate for a focused search for $\gamma$-ray pulsations.

Although all the available evidence points towards a redback classification, this needs to be confirmed with additional data. Ultimately this requires a detection of a pulsar in either radio or $\gamma$-ray observations. However, much deeper X-ray data than the shallow Swift/XRT observations presented here could allow the detection of orbital variability or a hard X-ray spectrum, which would provide compelling supporting evidence for our classification, and we were recently approved for \emph{XMM}-Newton observations for AO-20 (2021 May--2022 Apr).

The discovery of yet another redback candidate associated with a persistent $\gamma$-ray source that has been known since the first year after \emph{Fermi}'s launch suggests a substantial population of compact binaries still awaits detection, and that multi-wavelength follow-up of unassociated $\gamma$-ray sources remains a fruitful route to find new candidate MSPs.

\section*{Acknowledgements}

This research was performed while SJS held a NRC Research Associateship award at the Naval Research Laboratory. Work at the Naval Research Laboratory is supported by NASA DPR S-15633-Y.

We also acknowledge support
from NSF grant AST-1714825 and the Packard Foundation.

Based on observations obtained at the Southern Astrophysical Research (SOAR) telescope, which is a joint project of the Minist\'{e}rio da Ci\^{e}ncia, Tecnologia, Inova\c{c}\~{o}es e Comunica\c{c}\~{o}es (MCTIC) do Brasil, the U.S. National Optical Astronomy Observatory (NOAO), the University of North Carolina at Chapel Hill (UNC), and Michigan State University (MSU).

We acknowledge the use of public data from the Swift data archive.

\bibliography{main}

\end{document}